\documentclass{iopart}
\pdfoutput=1

\usepackage{iopams}

\usepackage{color}
\usepackage{dcolumn}
\usepackage{graphicx}
\usepackage[utf8]{inputenc}
\usepackage{latexsym}

\usepackage{amstext}

\newcommand{\scri}{\ensuremath{\mathcal{J}^+}}

\newcommand{\p}{\ensuremath{\partial}}

\definecolor{rgb_blue}{rgb}{0.0,0.0,0.6}
\definecolor{rgb_green}{rgb}{0.0, 0.6, 0.0}
\definecolor{rgb_red}{rgb}{0.7,0.,0.}

\newcommand{\eqref}[1]{Eq.~(\ref{#1})}

\newcommand{\sYlm}[2]{\ensuremath{{}_{#1}Y_{#2}}}

\newcommand{\sZlm}[2]{\ensuremath{{}_{#1}Z_{#2}}}

\newcommand{\norm}[1]{\ensuremath{\Arrowvert {#1} \Arrowvert}}

\renewcommand{\theenumi}{\arabic{enumi}.}

\renewcommand{\theenumii}{\roman{enumii}.}

\begin{document}

\title{General relativistic null-cone evolutions with a high-order scheme}

\author{Christian Reisswig\footnote{Einstein Fellow}}
\address{
  Theoretical Astrophysics Including Relativity,
  California Institute of Technology,
  Pasadena, CA 91125, USA
}

\author{Nigel Bishop}
\address{
  Department of Mathematics,
  Rhodes University,
  Grahamstown, 6139
  South Africa
}

\author{Denis Pollney}
\address{
  Department of Mathematics,
  Rhodes University,
  Grahamstown, 6139
  South Africa
}

\date{\today}

\begin{abstract}
We present a high-order scheme for solving the full non-linear
Einstein equations on characteristic null hypersurfaces 
using the framework established by
Bondi and Sachs. This formalism allows asymptotically flat spaces to be
represented on a finite, compactified grid, and is thus ideal for far-field
studies of gravitational radiation.
We have designed an algorithm
based on 4th-order radial integration and
finite differencing, and a spectral representation of angular components.
The scheme can offer significantly more accuracy with relatively low
computational cost compared to previous methods as a result of the higher-order
discretization. Based on a newly implemented code,
we show that the new numerical scheme remains stable and is
convergent at the expected order of accuracy.
\end{abstract}


\maketitle

\section{Introduction}
Characteristic formulations of the Einstein equations have proven to be an important tool for numerical
relativity. Most recently, they have been employed for the practical problem
of measuring gravitational waves in a gauge-invariant and unambiguous way from
numerically evolved spacetimes of binary black hole
mergers~\cite{Reisswig:2009us,Reisswig:2009rx,Babiuc:2010ze,
Babiuc:2011qi}, rotating stellar
core collapse~\cite{Reisswig:2010cd}, and collapsar formation~\cite{ott:11a}. 
The technique, called \emph{Cauchy-characteristic extraction} (CCE) (see
\cite{Winicour05} for a review)
takes metric boundary data produced by a 3+1
evolution on a worldtube $\Gamma$ of finite radius,
and uses null-cone evolutions of the Einstein equations to
transport
that metric data to future null infinity, $\mathcal{J}^{+}$,
the conformal outer boundary of
spacetime where gravitational radiation is invariantly defined and
interpreted in the Bondi gauge \cite{Bondi1962, Sachs1962, Penrose1963}.
Thus, this technique removes the influence of near-zone and
coordinate effects~\cite{Reisswig:2009us,Reisswig:2009rx,Babiuc:2010ze,
Babiuc:2011qi,Reisswig:2010cd}.

The null formulation is extremely efficient for evolving fields in the
wave zone, where the null coordinates are well-behaved and caustics along
geodesics are unlikely to be an issue. On relatively small computational grids, by
comparison with standard 3+1 methods, it is possible to achieve an
accuracy which has proven to be sufficient for practical applications
(gravitational wave measurement from compact bodies). In the case of
characteristic gravitational-wave extraction, the inner boundary data for
characteristic evolution is constructed on a worldtube at some distance
form the source where the curvature gradients are already rather small
compared to those close to the source. Thus, characteristic extraction
requires comparatively little numerical resolution and is therefore less
computationally demanding than 3+1 evolutions in the near-zone of a
dynamical source.

Although the computational effort is smaller, it is still
non-trivial. To yield sufficient accuracy, for example to extract
the gravitational radiation from a
binary black hole evolution, the characteristic computation by current
methods requires several days on up to a dozen processors on a workstation or small
cluster to complete a $3000M$ time-series encompassing a dozen orbits of
the binary (where $M$ is the dimensionless mass
of the spacetime).
The application of higher-order discretization schemes, on the other hand, may
deliver sufficient accuracy at much lower computational cost so that the
additional effort of characteristic
extraction would become negligible.
If the characteristic code could be run concurrently with the underlying 3+1
simulation, this would allow for on-the-fly extraction of waveforms, as
well as the interesting potential of using characteristic methods to provide
exact non-linear boundary conditions for 3+1 codes.

Characteristic codes have a long history in numerical relativity. A prominent
result was the first stable dynamical evolution of a black hole spacetime
in three spatial dimensions achieved by the Pittsburgh
group~\cite{Bishop96,Bishop97b}. Since then, 
the Pittsburgh null code (or \texttt{PITTNullCode}) has become the main building
block for current implementations of characteristic extraction used in numerical
relativity simulations
\cite{Reisswig:2009us,Reisswig:2009rx,Babiuc:2010ze,
Babiuc:2011qi,Reisswig:2010cd}, and is now part of the publicly available
Einstein Toolkit \cite{Loeffler12}. 
The code employs a single-null coordinate system, and is formulated in terms
of spin-weighted variables that are related to the original variables defined by
Bondi and collaborators \cite{Bondi1962, Sachs1962, Penrose1963}. It is built
around a numerical scheme based on points located at the corners of
null parallelograms, which was originally shown to be
stable~\cite{Gomez92b} in the context of the scalar wave equations.
The \texttt{PITTNullCode} in its original form is 2nd-order accurate in space
and time discretization, meaning that the error decreases as
$O(1/N^{2})$ as the resolution $N$ is increased.

Over the years, there have been several improvements to the original algorithm
as implemented in~\cite{Bishop96,Bishop97b}. Stereographic coordinates
were replaced by more uniform angular grids~\cite{Reisswig:2006, Gomez:2007cj}. Fourth-order accurate angular
derivatives have also been introduced,
though retaining the 2nd-order parallelogram-based radial integration
scheme~\cite{Reisswig:2006, Babiuc:2010ze}.

An alternative to the 2nd-order accurate null parallelogram was
attempted by Bartnik and Norton~\cite{Bartnik:1996js, Bartnik-Norton-1999}.
They developed an algorithm based on a null quasi-spherical
gauge, using 
method-of-lines integrators and 4th-order in time accuracy.
They also used a spherical
harmonic decomposition of variables on angular shells, and thus their code was
pseudo-spectral in the angular directions. Their formulation of the null
evolution equations (and coordinates) lead
to certain numerical complications, including the need for high-accuracy
interpolation operators to compute radial derivatives, and an elliptic gauge
equation. Ultimately, the code did not demonstrate long-term stability.

In this paper, we present a new high-order integration scheme for
characteristic evolutions in general relativity.
The evolution equations are written in
the Bondi form, following the prescription of~\cite{Bishop97b}. Time and
outward radial integrations are performed by method-of-lines schemes. In
particular, for the time integration, we use a classical 4th-order
Runge-Kutta integrator.
For the radial direction, 
we use a modified Adams-Moulton multi-step method. A multistep method
 is required by the lack
of information at points between grid spacings, which is needed by the intermediate steps of the Runge-Kutta schemes.
A similar method, for the
case of 2nd-order accuracy and axisymmetry, was previously used
in~\cite{Bishop90}. 
In the present context of full three-dimensional characteristic evolutions,
we additionally need to discretize the angular direction. For this, we use
spectral expansions in terms of spin-weighted real-valued spherical harmonics.
The equations are solved on an angular collocation grid using the
pseudo-spectral method. Radial and time integrations are performed on
the angular spectral coefficients of the evolution variables.

By referring to a simplified linear model problem in
Section~\ref{s-stability}, we argue analytically
that the proposed method is stable.
We review
the Bondi evolution equations in Section~\ref{sec:einstein-eqns}. In
Section~\ref{sec:num-methods}, we describe the numerical methods, including
time and radial integrators, and pseudo-spectral derivatives.
Finally, we test a newly implemented three-dimensional high-order
code, demonstrating the expected order of convergence and accuracy of the
method.
We find that the new scheme is significantly more efficient than the old scheme,
reaching the same level of accuracy using only very few radial and angular
points.

\section{Integration schemes and stability}
\label{s-stability}

Characteristic evolution schemes based on null or double-null coordinates
have a significantly different character than conventional 3+1 (Cauchy) evolutions.
In 3+1 evolutions, spacetime
is foliated along a timelike vector field $t^\alpha$ by spacelike hypersurfaces $\Sigma$. 
In so-called 2+1+1 characteristic evolutions (the characteristic initial
boundary value problem~\cite{Stewart:1990uf}), spacetime is
foliated along a timelike vector field by null hypersurfaces,
which are characteristic surfaces of the Einstein field
equations.
In this section, we describe a simplified model characteristic problem which exhibits the important features of the Einstein system, which we
outline explicitly in Section~\ref{sec:einstein-eqns}.

The characteristic method solves for the values of a field, $J$,
which obeys a hyperbolic equation of the form
\begin{equation} \label{eq:generic-evol}
    J_{,ur} = F(J, J_{,r}, J_{,rr}, J_{,A})\,.
\end{equation}
Here $u=t-r$ is a retarded time coordinate labeling individual
null slices, $r$ is a
radial coordinate along each null slice, and the index $A=(\theta,\phi)$
labels angular coordinates. We allow the function $F$ to depend on $J$,
and its partial derivatives (to 2nd-order in $r$), which we label
by subscripts
\begin{equation}
    \frac{\p^2 J}{\p u \p r} \;\Rightarrow\; J_{,ur}\,.
\end{equation}
We consider the problem on a domain bounded on the interior by a
timelike worldtube, $\Gamma$, at some finite areal radius
$R=R(\theta,\phi)$ from
the centre of our coordinate system, and on the exterior by $\scri$.
Boundary data consists of the variables required to evaluate $F$ on
$\Gamma$, as well as initial data for $J$ along a single null
cone at $u=u_{0}$.

We introduce an intermediate variable $\Phi = J_{,u}$, which allows
us to recast~\eqref{eq:generic-evol} as a pair of 1st-order equations
\begin{numparts}
    \begin{eqnarray}
        \Phi_{,r} &=& F(J, J_{,r}, J_{,rr}, J_{,A})\,, \label{eq:generic-phi} \\
        J_{,u} &=& \Phi\,.  \label{eq:generic-J}
    \end{eqnarray}
    \label{eq:simple-null}
\end{numparts}
Equation~(\ref{eq:generic-phi}) does not involve time derivatives $J_{,u}$
and can thus be solved on a $u=\text{constant}$ null slice by radial
integration. Given data for $J$ on a $u=\text{constant}$ slice either from
the last time step or by appropriate initial data, we  propagate the
system forward in time by first solving~\eqref{eq:generic-phi} along
the radial direction, and then integrating~\eqref{eq:generic-J}
forward in time to determine $J$ on the next slice.

The schemes which we use for both radial and null integrations
fall within the broad class of method-of-lines integrators for partial
differential equations. These schemes assume that we have been able to
evaluate the RHS at a point, so that standard ordinary differential
equation (ODE) solvers can be applied to evolve the functions either
forward in time, $u$, or (in the case of $\Phi$) outward in $r$.

For the time direction, we can use an explicit integrator, such as a
standard 4th-order Runge-Kutta scheme.  Recall, however, that in taking
the solution from $u$ to $u+\Delta u$, the
Runge-Kutta method involves calculating the right-hand side (RHS)
at a number of intermediate steps. 
In the case of the $J$ integration the RHS is $\Phi$, and thus we need
to compute $\Phi$ via radial integration of~\eqref{eq:generic-phi} at
intermediate substeps between our timesteps of size $\Delta u$.

Applying such a scheme in the radial direction for $\Phi$ is more
problematic. In that case, the RHS is given by $F$, which is a function of
data which is only known on discrete spheres (separated by a fixed
distance, $\Delta r$), and so we cannot evaluate the intermediate substeps
required by a Runge-Kutta-type integrator. Alternatively, we could make
use of a multi-step algorithm, such as Adams-Bashforth or Adams-Moulton.
These methods evaluate the RHS over some number of previous points, which in
the case of the radial integration correspond to a set of the radial
spheres on which $J$ and its derivatives can been evaluated.

We have examined a number of numerical schemes for carrying out the radial
integration required by~\eqref{eq:generic-phi} for the Einstein system,
but almost universally found them to be unstable in empirical tests. To
investigate the stability of different numerical methods, we turn to
a simpler model which embodies the main features of the Einstein system,
on which we can carry out a von Neumann analysis.

By setting individual terms in the Einstein equations to zero and
examining the subsequent numerical evolution, we came to the conclusion
that the key terms determining the stability are those which involve the
variables $J_{,r}$ and $J_{,rr}$. Thus we constructed two simplified
systems which consist exclusively of these terms in the radial integration
of a variable $\Phi=J_{,u}$. That is, we considered the systems
\begin{equation}
    \Phi_{,r} = J_{,r}\,,\qquad J_{,u} = \Phi\,,{}
    \label{e-phistabr}
\end{equation}
and
\begin{equation}
    \Phi_{,r} = J_{,rr}\,,\qquad J_{,u} = \Phi\,,{}
    \label{e-phistabrr}
\end{equation}
individually. In practice, we replace the areal radius $r$ by a compactified coordinate $x$, defined by
\begin{equation}
    x = \frac{1}{r_{\Gamma} + r}
\end{equation}
where $r_{\Gamma}$ is a non-zero parameter corresponding to the radius
of the inner boundary (see~\eqref{e-compactify}, below), and have found
this transformation to be important in the stability analysis. In terms
of $x$, Eqs. (\ref{e-phistabr}) and (\ref{e-phistabrr}) become
\begin{equation}
    \Phi_{,x} +\frac{(r+r_\Gamma)^2}{r\,r_\Gamma}\Phi
         = \frac{r_\Gamma}{r(r+r_\Gamma)}J_{,x}\,,\label{e-phstabx}\qquad
    J_{,u} = \Phi \label{e-Jstab}\,,
\end{equation}
and
\begin{equation}
    \Phi_{,x} +\frac{(r+r_\Gamma)^2}{r\,r_\Gamma}\Phi
        =  \frac{r_\Gamma}{2(r+r_\Gamma)^2}J_{,xx}\,,
        \label{e-phstabxx}\qquad
    J_{,u} = \Phi\,,
\end{equation}
respectively.

The von Neumann analysis corresponds to assuming the following form for
the model variables:
\begin{numparts}
    \begin{equation}
        J=e^{wu}e^{ikx}\,,\qquad \Phi=E J\,,{}
        \label{e-vN-1}
    \end{equation}
    \begin{equation}
        \Delta x=h\,,\qquad \Delta u= \mu h\,,\qquad \nu=\frac{kh}{2}.
        \label{e-vN-2}
    \end{equation}
\end{numparts}
The $J_{,u}$ equations are evolved using a 4th-order Runge-Kutta integration,
for which the
general stability analysis is quite involved. However, to leading order in
$h$, it is the same as the Euler method (and this was also found for other
evolution algorithms such as the Adams-Bashforth methods). Thus,
\begin{equation}
    e^{w\mu h}=1+\mu h E,
\end{equation}
and the stability is determined by the sign of $\Re(E)$:
\begin{eqnarray}
\Re(E) \;\cases{ > 0, & \text{Unstable}, \\
       = 0:& \text{Stability unknown}, \\
       < 0:& \text{Stable.}
       }
\end{eqnarray}
The quantity $E$ is determined by the particular finite difference
algorithm used to evaluate~\eqref{e-phstabx} or~\eqref{e-phstabxx},
followed by substitution of Eqs. (\ref{e-vN-1}, \ref{e-vN-2}).
We investigated both second and 4th-order explicit
(Adams-Bashforth) and implicit (Adams-Moulton) multi-step methods, with
finite differences evaluated using centred, forward and backward methods
of the appropriate accuracy. The calculations are somewhat lengthy and
were done using
a computer algebra script (see the file \texttt{vN\_comp3.map}
in the online supplement).

We found that the case~\eqref{e-phstabx} involving $J_{,x}$ always leads to
values of $\nu$ for which $\Re (E)>0$ so that the
system is always unstable, independent of the integration scheme for
$\Phi$. However, there are cases for which~\eqref{e-phstabxx},
involving $J_{,xx}$, is stable.
Since a second derivative term in the stability analysis is divided
by $h^2$ compared to division by $h$ for the first derivative term,
to leading order in $h$ the stability of the second derivative term
dominates that of the first derivative term.

The stability analysis
of~\eqref{e-phstabxx} indicates that among the tested methods, only
the following cases are stable:
\begin{itemize}
\item 2nd-order, Adams-Moulton, centred differences;
\item 2nd-order, Adams-Bashforth, forward differences;
\item 2nd-order, Adams-Moulton, forward differences;
\item 4th-order Adams-Moulton, forward differences;
\end{itemize}
where by ``forward'' difference operator we indicate that radial derivatives
of $J$ are calculated using a stencil which involves points in the
positive radial direction. We were able to confirm these results by
empirical tests with the simplified system,~\eqref{e-phstabxx}. Thus,
for the full Einstein equations, described in the next section, we
implemented a scheme in which radial integrations are carried out using
a 4th-order Adams-Moulton method with upwinded radial derivatives.

\section{The Einstein equations in the Bondi-Sachs framework}
\label{sec:einstein-eqns}

\subsection{Coordinates} The Bondi formulation writes the Einstein
equations in terms of a null foliation of an asymptotically flat
spacetime. We introduce coordinates $y^\alpha=(r,y^A,u)$. The coordinate
$r$ is a radial surface area coordinate, and $u=t-r$ is a retarded time
coordinate which replaces the time, $t$, of 3+1 formulations. The $y^{A}$
are angular coordinates, labeled by uppercase indices that take the
values $1$ and $2$. The coordinates $(y^{A}, u)$ label individual null
geodesics extending from a  worldtube $\Gamma=S^{2}\times\mathbb{R}$. The
worldtube is chosen so that $r=\text{constant}$ on $\Gamma$. In
these coordinates, the general spacetime line element is
\begin{eqnarray}
    ds^2 &=& -\Big(\e^{2\beta}\frac{V}{r}-r^2h_{AB}U^AU^B\Big)\,du^2
    - 2\e^{2\beta}\,du\,dr \nonumber \\
    & &\qquad{}
    -2r^2h_{AB}U^B\,du\,dy^A+r^2h_{AB}\,dy^A\,dy^B\,,{}
    \label{eq:Bondi-metric}
\end{eqnarray}
where $h_{AB}$ satisfies
\begin{equation}
    h^{AB}h_{BC}={\delta^A}_C, \qquad \det(h_{AB})=\det(q_{AB})\,,{}
    \label{eq:null1}
\end{equation}
and $q_{AB}$ is the unit sphere metric. 

The spacetime is described by
$V$, $\beta$, $U^{A}$, and $h_{AB}$, which are functions of the coordinates.
It is convenient to write quantities in terms of spin-weighted scalars in
order to remove explicit angular tensor components. This simplifies the
expression of the field equations in a way that is independent of the
choice of angular coordinates. To this end, we introduce a complex dyad $q^A$
satisfying $q^{A} = q^{AB}\,q_{B}$, $q^{A}\,q_{B}=0$, $q^{A}\bar{q}_{A}=2$.
\footnote{An explicit form for $q^A$ will not be needed here,
since we will represent angular dependence in terms of spin-weighted
spherical harmonic basis functions constructed using a particular dyad
representation adapted to the choice of angular coordinates.}. By
projecting the angular variables onto this dyad, we define the complex
valued scalars
\begin{equation}
    J = \frac{1}{2}\,h_{AB}q^{A}q^{B}\,,\qquad
    K = \frac{1}{2}\,h_{AB}q^{A}\bar{q}^{B}\,,\qquad
    U = U^{A}q_{A}\,,
\end{equation}
of spin weights 2, 0, and 1, respectively. The components of
$h_{AB}$ are uniquely determined by $J$ due to the determinant condition,
\eqref{eq:null1}, thus fixing $K$ as a function of $J$ via 
\begin{equation} \label{eq:K}
K = \sqrt{1+J\bar{J}}\,.
\end{equation}
Corresponding to the complex dyad, we introduce complex angular covariant
differential operators $\eth$ and $\bar{\eth}$ which maintain the property
of spin-weight when acting on a scalar $\Phi$ of spin-weight $s$~\cite
{Gomez97}. The action of the $\eth$ and $\bar\eth$ operators is
restricted to transformations of our spin-weighted spherical harmonic
basis functions (see Section~\ref {sec:ang-derivs}).

Before writing out the Einstein equations, we note that it is convenient
to introduce an additional intermediate variable defined by
\begin{equation}
    Q := r^2 e^{-2\,\beta} h_{AB} U^B_{,r} q^A\,.
\end{equation}
This spin-weight 1 variable, which is the first radial derivatives of $U$,
will allow us to write the equations in 1st-order form. Also, we
re-express $V$ in terms of a new variable
\begin{equation}
    \hat{W} := \frac{V - r}{r^2}\,,
\end{equation}
which has a regular limit as $r\rightarrow\infty$.

\subsection{Einstein equations}
In Bondi coordinates, the vacuum Einstein equations
\begin{equation}
    R_{ab} = 0\,
\end{equation}
give rise to a hierarchy of equations which we can characterize as (i)
hypersurface equations, (ii) evolution equations, and (iii) constraints.

Hypersurface equations do not depend on $u$-derivatives, and thus can be
evaluated within a $u=\text{constant}$ slice. They are determined by the
components $R_{rr}$, $R_{rA}q^{A}$, and $R_{AB}h^{AB}$
and lead to the following hierarchy of equations:
\begin{numparts}
\begin{eqnarray}
    \beta_{,r} &=& N_\beta\,,{}
        \label{eq:beta} \\
    (r^2 Q)_{,r} &=& -r^2 (\bar \eth J + \eth K)_{,r}
        +2r^4\eth \left(r^{-2}\beta\right)_{,r} + N_Q\,,{}
            \label{eq:wq} \\
     U_{,r} &=& r^{-2}e^{2\beta}Q +N_U\,,{}
        \label{eq:wua} \\
    (r^{2}\hat{W})_{,r} &=& \frac{1}{2} e^{2\beta}{\cal R} -1
        - e^{\beta} \eth \bar \eth\, e^{\beta} \nonumber \\
        & &+ \frac{1}{4} r^{-2} \big(r^4
        (\eth \bar U +\bar \eth U )
        \big)_{,r} + N_W\,,                            \label{eq:ww}
\end{eqnarray}
\end{numparts}
where the Ricci scalar is given explicitly by
\begin{equation}
    {\cal R} =2 K - \eth \bar \eth K + \frac{1}{2}(\bar \eth^2 J
        + \eth^2 \bar J)
        +\frac{1}{4K}(\bar \eth \bar J \eth J
            - \bar \eth J \eth \bar J)\,,
    \label{eq:calR}
\end{equation}
and $N_\beta$, $N_Q$, $N_U$, and $N_W$ are non-linear aspherical terms
given explicitly in~\ref{app:nl}. The equations are solved in
succession, assuming
available data $J$ on a $u=\text{constant}$ slice and constraint satisfying
inner boundary data at the worldtube $\Gamma$ for each of the hypersurface
variables. 
This allows us to solve
for $\beta$, which in turn provides data for the equation for $Q$. 
Given $\beta$ and $Q$, we can then solve for $U$, and finally for $\hat{W}$.

The $R_{AB}q^{A}q^{B}$ component of the Einstein equations determines
the evolution equation for $J$:
\begin{eqnarray}
    && 2 \left(rJ\right)_{,ur}
    - \left(\big(1+r \hat{W}\big)\left(rJ\right)_{,r}\right)_{,r} = \nonumber \\
    && \qquad -r^{-1} \left(r^2\eth U\right)_{,r}
    + 2 r^{-1} e^{\beta} \eth^2 e^{\beta}- \big(r \hat{W} \big)_{,r} J
    + N_J\,,
    \label{eq:wev}
\end{eqnarray}
where the non-linear aspherical terms have been gathered in the quantity
$N_J$ (specified in~\ref{app:nl}). We 
introduce an intermediate variable
\begin{equation} \label{eq:def-phi}
    \Phi:=J_{,u}\,.
\end{equation}
In terms of $\Phi$, ~\eqref{eq:wev} becomes a new hypersurface equation
\begin{eqnarray}
    && 2 \left(r\Phi\right)_{,r}
    - \left(\big(1+r \hat{W}\big)\left(rJ\right)_{,r}\right)_{,r} = \nonumber \\
    && \qquad -r^{-1} \left(r^2\eth U\right)_{,r}
    + 2 r^{-1} e^{\beta} \eth^2 e^{\beta}- \big(r \hat{W} \big)_{,r} J
    + N_J\,.
    \label{eq:phi}
\end{eqnarray}
which is integrated radially from $\Gamma$ using known values of the
hypersurface variables determined in~\eqref{eq:beta}-(\ref{eq:ww}).
Then, $J$ is determined by a timelike integration of
\begin{equation}
    J_{,u} = \Phi\,.
\label{eq:Ju}
\end{equation}
We have expressed the Einstein system in a form analogous to the
simplified model described in~\eqref{eq:simple-null}. The source for $\Phi$
is complicated, but determined entirely by radial integration. Note the
presence of the $J_{,rr}$ in the second term of \eqref{eq:phi}, which we
have highlighted in Section~\ref{s-stability} as key to determining the
stability of numerical evolution schemes.

Finally, we take advantage of the nature of null geodesics in
asymptotically flat spacetimes to compactify the radial direction so that
$\scri$ is a boundary point of a closed domain. We replace the areal
radius $r$ by a new coordinate $x$ via the invertible coordinate
transformation
\begin{equation}
    x(r)= \frac{r}{r_{\Gamma}+r}\,,\qquad r(x)=r_{\Gamma}\frac{x}{1-x}\,,
\label{e-compactify}
\end{equation}
where $r_{\Gamma}$ is a constant, which we choose to be the radius of
the worldtube, $\Gamma$. In this coordinate, the equations have a regular
limit as $x\rightarrow 1$, and furthermore, we are able to set terms of order
$1/r^{n}$ for $n=1,2,\ldots$, to zero at $\scri$ (see Section~\ref{sec:scri}).
Throughout the domain, derivatives are evaluated numerically in terms of
the new coordinate, $x$, and then transformed into $r$-derivatives using
the standard Jacobian transformations. For \eqref{e-compactify}, these
are
\begin{equation} \label{eq:jac}
\frac{d x}{d r}=\frac{r_\Gamma}{(r+r_\Gamma)^2}
, \qquad{} \frac{d^2 x}{d r^2}=-\frac{2r_\Gamma}{(r+r_\Gamma)^3}.
\end{equation}

\subsection{Form of the equations at $\scri$}
\label{sec:scri}

In the limit of $r\rightarrow\infty$, corresponding to $x=1$, the
numerical treatment of the equations requires special care.
The problematic terms are those involving the Jacobian and the coordinate
function $r(x)$, which are not regular as $r\rightarrow\infty$.
If the coordinate function and the Jacobian are explicitly inserted into the
equations, using their form given by \eqref{e-compactify} and
\eqref{eq:jac}, respectively, divergent terms are seen to cancel. However, 
since we do not explicitly impose a specific compactification --- we have formulated the problem in terms of generic Jacobians rather than the
specific formulas of Eqs.~(\ref{e-compactify}) and (\ref{eq:jac}) --- we need to be careful
to avoid irregular terms at~$\scri$.

It is sufficient to require that in the limit $r\rightarrow\infty$,{}
the compactified coordinate transformation and its Jacobian 
approach the explicit forms given in \eqref{eq:jac} and
\eqref{e-compactify}, respectively.
For instance, consider the equation 
\begin{equation}
 U_{,x}=\Big(r^2(x)\frac{d x}{d r}\Big)^{-1}\e^{2\beta}\Big(Q+r^2 e^{-2\beta} N_U\Big)\,.{}
\end{equation}
According to \eqref{eq:jac} and
\eqref{e-compactify}, we have
\begin{equation}
\Big(r^2(x)\frac{d x}{d r}\Big)^{-1} \;\rightarrow \;r_\Gamma^{-1}
\end{equation}
as $r\rightarrow\infty$. Hence, at $\scri$,
\begin{equation}
 U_{,x}=r_\Gamma^{-1}\e^{2\beta}\Big(Q+r^2 e^{-2\beta} N_U\Big)
=r_\Gamma^{-1}\e^{2\beta}\Big(KQ-J\bar{Q} \Big)\,.{}
\end{equation}
We proceed in a similar manner for the other hypersurface equations.
The specific form of the equations at $\scri$ is given in \ref{app:reg-scri}.
Note additionally that for $Q$, $\hat{W}$ and $\Phi$ it is possible to directly
evaluate the respective quantity without radial integration at $\scri$.
For instance, as $r\rightarrow\infty$,
\begin{equation}
Q = -2 \eth \beta\,.
\end{equation}

\section{Numerical methods}
\label{sec:num-methods}

\subsection{Discrete representation of the evolution variables}
\label{sec:discrete-repr}

The evolution algorithm is a hybrid of finite-difference (for radial
and time integration) and pseudo-spectral (for angular directions)
methods. In the compactified radial direction $x$, fields $\Phi$ are evaluated
on a uniform
grid of $N_x$ points, $\Phi_{i}$, $i=0,\ldots, N_x-1$ at points
$x\in[x_{\rm in},\ldots,1]$.
The inner coordinate radius, $x_{\rm in}=x_{i=0}$, is that of the world-tube
$\Gamma$, where we need to specify
appropriate boundary data 
at any given time $u$ to carry out a radially outward hypersurface
integration. As we will see in Section~\ref{sec:rad-int}, our radial
integration
scheme
actually requires $3$ radial points to start
the algorithm.
We therefore need to provide boundary data on the first
$i=0,1,2$ radial points so that our worldtube $\Gamma$ spans three radial
points.
Boundary data is required for
\begin{equation}
    \left\{\beta, Q, U, \hat{W}, \Phi\right\}\quad\text{for all}\quad
    y^{A}_{i=0,1,2}\in\Gamma|_{u}\,.
\end{equation}
The outer boundary of the compactified radial grid is placed at the
outermost gridpoint $i=N_x-1$ corresponding to future null infinity $\scri$.

At each radial point $x_i$, we represent angular dependence as a spectral
expansion in terms of real-valued spin-weighted spherical harmonics, according
to
\begin{equation} \label{eq:spec-epand}
    \Phi_i(y^{A}) = \sum_{\ell=s}^\infty \sum_{m=-\ell}^{m=+\ell}
        \Phi_{\ell m}(x_i)\;{}_sZ_{\ell m}(y^{A})\,,
\end{equation}
where the $\sZlm{s}{\ell m}$ are spin $s$ real-valued spherical harmonics,
defined in terms of the standard $\sYlm{s}{\ell m}$ basis \cite{Goldberg:1967}
by
\begin{equation}
    \sZlm{s}{\ell m} = \cases{
            \frac{1}{\sqrt{2}}\left(\sYlm{s}{\ell m}
                + (-1)^{m}\sYlm{s}{\ell -m}\right)\,, & $m> 0$\,,{}
                \\
            \sYlm{s}{\ell m}\,, & m=0\,,\\
            \frac{i}{\sqrt{2}}\left((-1)^{m}\sYlm{s}{\ell m}
                - \sYlm{s}{\ell -m}\right)\,, & $m< 0$\,. \\
            }
\end{equation}
We use the $\sZlm{s}{\ell m}$ to accommodate the real-valued spin-0
quantities, which naturally yield real valued coefficients.

We store a finite number of harmonic coefficients for each variable,
terminating the sum according to the maximum number of measurable
gravitational wave modes contained in the solution. For our test case with
linearized solutions as discussed in Section~\ref{sec:results}, this is
$\ell_{\rm
max}=3$. For the realistic case of current binary black hole merger simulations
the number of resolved gravitational-wave modes in the 3+1 evolution is 
typically $\ell_{\rm max}\sim8$, beyond which their amplitude is below the
level of discretization error. Although this case is not considered here and is
left for future work, we do present results of a stability test in which
$\ell_{\rm max}=8$.
We store the spectral coefficients of the expansion
of each evolution variable at each point of the radial grid.

Radial and time integration is performed entirely on the spherical
harmonic coefficients of the evolution variables, with the one exception
being \eqref{eq:phi2}, to be introduced below.
This equation contains non-linear terms of the form $a\,\Phi$ where $a$ and
$\Phi$ are both functions of angular coordinates $y^A$.
To compute
non-linear terms occurring either in \eqref{eq:phi2} or in the RHS for a given
hypersurface equation, we first need
to recompose the involved variables on a collocation grid. After the terms
have been evaluated on the
collocation grid, we decompose them back into real-valued spin-weighted
spherical harmonics.

We construct a pseudo-spectral collocation grid for spherical harmonics 
by defining a set of grid points on a spherical shell $S^2$ using a
$y^{A}=(\theta,\phi)$ spherical-polar coordinate system with
constant grid spacing in $\theta$ and $\phi$ direction
\begin{equation} \label{eq:collocation-points}
\hspace{-1cm}\left\{ (\theta_j, \phi_k)=\left(\pi
\frac{j+\frac{1}{2}}{N_\theta},2\pi\frac{k}{N_\phi}\right)
\,:\, j,k\in\mathbb{N};\, 0\leq j<N_\theta,\, 0\leq k < N_\phi\right\}\,.
\end{equation}
Recomposing quantities on the collocation grid is easily done by evaluating
the sum of the spherical harmonic expansion via
\begin{equation}
f(\theta_j,\phi_k) = \sum_{\ell m} f_{\ell m} {}_s Z_{\ell
m}(\theta_j,\phi_k)\,.
\end{equation}

Decomposing a quantity into spherical harmonics requires surface
integration over $S^2$. The expansion coefficients are computed according to a
discrete version of
\begin{equation} \label{eq:sZlm-decomp}
f_{\ell m} = \int_\Omega d\Omega f(\theta,\phi) {}_s \bar{Z}_{\ell m}(\theta,
\phi)\,,
\end{equation}
where $d\Omega=r^2\sin\theta \,d\theta \,d\phi$ is the surface element on the
collocation grid.
A numerical integration algorithm which is exact for spherical harmonics
up to order $(\ell,m)$ is given by Gauss-Chebyshev quadratures using
$(N_\theta,N_\phi)=(2(\ell+1),2 (\ell+1))$ points on $S^2$ (
e.g.~\cite{Press02}).
This algorithm makes use of coordinate dependent weights $w_j$ in $\theta$
direction.
In $\phi$, the weights are simply $1$ since $\phi$ is a periodic coordinate
direction.
Since we use equally spaced points in $\theta$ (equivalent to Chebyshev nodes in
$x\equiv \cos\theta$), the weights in $\theta$ direction are
\cite{Driscoll94}
\begin{equation}
w_j =
\frac{4}{N_\theta}\sum_{\ell=0}^{N_\theta/2-1}\frac{1}{2\ell+1}
\sin((2\ell+1)\theta_j)\, .
\end{equation}
The integral \eqref{eq:sZlm-decomp} reduces to
\begin{equation}
f_{\ell m} =
\frac{\pi}{N_\theta}\frac{2\pi}{N_\phi}\sum_{j=0}^{N_\theta}\sum_{k=0}^{N_\phi}
f(\theta_j,\phi_k) {}_s \bar{Z}_{\ell m}(\theta_j,\phi_k) \sin\theta_j\, w_j\,,
\end{equation}
provided we have 
\begin{equation}
(N_\theta,N_\phi)\leq(2(\ell+1),2(\ell+1))
\end{equation}
points on $S^2$.

We have thus established an exact mapping between the representation in
terms of spherical harmonic coefficients and the representation on the
collocation grid. To speed up the computation, we precompute the ${}_s
Z_{\ell m}$, and the product $w_j \sin\theta_j$.

As an example for our algorithm, consider the linearized version of the
hypersurface equation for $U$,
\begin{equation}\label{eq:hyperU}
    U_{,x}=\left(r^2(x)\frac{d x}{d r}\right)^{-1}\e^{2\beta}Q\,.{}
\end{equation}
We recompose $\beta$ and $Q$ on the sphere using their spectral
expansion coefficients in order to carry out the required multiplications. The
complete procedure for integrating~\eqref{eq:hyperU} can be summarized as
follows.
\begin{enumerate}
    
     \item Loop over radial points. On each radial shell $x_i$:
     \begin{enumerate}
          \item
            Define grid points on a spherical shell $S^2$ according to
            \eqref{eq:collocation-points}.        
          \item
            Recompose
            \begin{equation}
                \beta(x_i,\theta_j,\phi_k)=\sum_{\ell m} \beta^{\ell m}(x_i)\;
                    \sZlm{0}{\ell m}(\theta_j,\phi_k)\,,{}
            \end{equation}
            and
            \begin{equation}
                Q(x_i,\theta_j,\phi_k)=\sum_{\ell m} Q^{\ell m}(x_i)\;
                    \sZlm{1}{\ell m}(\theta_j,\phi_k)\,.
            \end{equation}
        
          \item
            Loop over all angular points $\theta_j$ and $\phi_k$. For each
            angular point, compute $U_{,x}(\theta_j,\phi_k)$ using
            ~\eqref{eq:hyperU}.
            
          \item \label{step:recompose}
            Decompose $U_{,x}(\theta_j,\phi_k)$ via
            \begin{equation}
                \hspace{-2cm}(U_{\ell m})_{,x}(x_i) =
\frac{\pi}{N_\theta}\frac{2\pi}{N_\phi}\sum_{j=0}^{N_\theta}\sum_{k=0}^{N_\phi}
U_{,x}(\theta_j,\phi_k) {}_s \bar{Z}_{\ell m}(\theta_j,\phi_k) \sin\theta_j\,
w_j\,.
            \end{equation}
    \end{enumerate}
    \item Radially integrate $(U_{\ell m})_{,x}(x_i)$ to obtain
        $U_{\ell m}(x_i)$.
\end{enumerate}
The last step, radial integration, is described in more detail in
Section~\ref{sec:rad-int}.

\subsection{Radial derivatives and dissipation}
\label{sec:rad-deriv}

Radial derivatives of all hypersurface quantities are generally obtained from the RHS
of their corresponding radial ODE integrations and hence do not need to be recomputed by means
of finite difference operators.
However, the metric variable $J$ (and also $K$) itself is not directly
obtained via radial integration and hence must
be computed everywhere.
We approximate $J_{,x}$ and $J_{,xx}$ by means of finite difference 
operators of 4th-order. The radial derivative of $K$ can be obtained by using
\eqref{eq:K}.

According to stability analysis and empirical findings
(see Section~\ref{s-stability}), we
apply fully side-winded derivatives with the stencil points in the
direction of $\scri$.
We use 4th-order first and second derivatives
\begin{eqnarray} 
\hspace{-1cm}\p f_i &=& \frac{1}{\Delta x}\left(-\frac{25}{12}f_i 
         + 4 f_{i+1} - 3 f_{i+2} + \frac{4}{3}f_{i+3} 
         - \frac{1}{4}f_{i+4}\right)\,, \label{eq:rad-deriv1} \\
\hspace{-1cm}\p^2 f_i &=& \frac{1}{\Delta x^2}\left(\frac{15}{4}f_i 
         - \frac{77}{6}f_{i+1} + \frac{107}{6}f_{i+2} - 13 f_{i+3} 
         + \frac{61}{12}f_{i+4} - \frac{5}{6}f_{i+5}\right),
\label{eq:rad-deriv2}
\end{eqnarray}
where $\Delta x$ is the grid spacing in the compactified radial coordinate
direction.

Close to $\scri$ when $i > N_x-5$, we switch to 4th-order centred stencils
\begin{eqnarray} 
\hspace{-1cm}\p f_i &=& \frac{1}{\Delta x}\left(+\frac{1}{12}f_{i-2} -
\frac{2}{3} f_{i-2} 
                                   +\frac{2}{3}f_{i+1} -
\frac{1}{12}f_{i+2}\right)\,, \label{eq:rad-deriv-bdry1} \\
\hspace{-1cm}\p^2 f_i &=& \frac{1}{\Delta x^2}\left(-\frac{1}{12}f_{i-2} +
\frac{4}{3}f_{i-1}
                                       -\frac{5}{2}f_{i} + \frac{4}{3} f_{i+1} -
\frac{1}{12}f_{i+2}\right), \label{eq:rad-deriv-bdry2}
\end{eqnarray}
and when $i > N_x-3$, we switch to side-winded stencils pointing towards the inner boundary
\begin{eqnarray} 
\hspace{-1cm}\p f_i &=& \frac{1}{\Delta x}\left(\frac{25}{12}f_i - 4 f_{i-1} 
        + 3 f_{i-2} - \frac{4}{3}f_{i-3} 
        + \frac{1}{4}f_{i-4}\right)\,, \label{eq:rad-deriv-bdryA1} \\
\hspace{-1cm}\p^2 f_i &=& \frac{1}{\Delta x^2}\left(\frac{15}{4}f_i -
\frac{77}{6}f_{i-1} 
        + \frac{107}{6}f_{i-2} - 13 f_{i-3} + \frac{61}{12}f_{i-4} 
        - \frac{5}{6}f_{i-5}\right). \label{eq:rad-deriv-bdrA2}
\end{eqnarray}

In addition, we apply a numerical dissipation operator to $J$.
We use a 5th-order Kreiss-Oliger dissipation operator of the form
\begin{equation} \label{eq:diss}
\hspace{-1cm}D f_i = \frac{\epsilon_{\rm diss}}{64 \Delta x} \left(f_{i-3} - 6
f_{i-2} + 15
f_{i-1} - 20 f_{i} + 15 f_{i+1} - 6 f_{i+2} + f_{i+3}\right),
\end{equation}
where $\epsilon_{\rm diss}$ controls the strength of the applied dissipation
operator $D$.
At the outer boundary (at $\scri$), where we do not have enough points to
compute the dissipation operator, we use one-sided stencil derived
for an overall 4th-order accurate summation-by-parts (SBP) operator
(though we do not make explicit use of the SBP property).
The particular stencil coefficients are derived in \cite{Diener05b1}. We
explicitly state the stencil coefficients in \ref{app:diss}.

Empirical tests have shown that radial dissipation applied to $J$ is
crucial to improve the stability properties of our scheme.

\subsection{Angular derivatives}
\label{sec:ang-derivs}

Numerical derivatives in the angular direction are obtained via analytic
angular derivatives of the spin-weighted real-valued spherical harmonic
spectral basis functions.
The action of the $\eth$ derivative on the real-valued spherical harmonics is
given by \cite{Goldberg:1967}
\begin{equation}
\eth \; {}_s Z_{\ell m} = \sqrt{(\ell+s+1)(\ell-s)}\;
    {}_{s+1}Z_{\ell m}\, .
\end{equation}
The action of $\eth$ on a spectrally expanded function is
\begin{equation}
\eth f = \sum_{\ell m} f_{\ell m} \sqrt{(\ell+s+1)(\ell-s)} \; {}_{s+1}Z_{\ell
m}\,.
\end{equation}
Similarly, the action of $\bar\eth$ is given by
\begin{equation}
\bar\eth \; {}_s Z_{\ell m} = -\sqrt{(\ell-s+1)(\ell+s)}\;  {}_{s-1}Z_{\ell
m}\,.
\end{equation}

\subsection{Radial integration}
\label{sec:rad-int}

The hypersurface equations are integrated in the radial
direction using a multistep method. The classes of methods that
we have studied for this problem are either the explicit Adams-Bashforth
methods, the implicit Adams-Moulton methods, and a combination of
the two in the form of a predictor-corrector scheme. However, as discussed
in Section~\ref{s-stability}, at 4th-order, explicit methods are unstable for
our particular set of equations.

Thus, the radial integration uses a fully implicit
method. Fortunately, the Einstein equations in Bondi-Sachs form are particularly
convenient for this purpose, as they form a hierarchy (as outlined in
Section~\ref{sec:einstein-eqns}) with only one unknown function in each equation. Furthermore, the equations are linear in this unknown.
Schematically, we write each equation in the form
\begin{equation}
\frac{dy}{dx}+y g(x)=f(x),
\end{equation}
and the 4th-order fully implicit Adams-Moulton scheme can be written in
explicit form as
\begin{eqnarray} \label{eq:rad-scheme}
y_{i+1}&\Big(1+\frac 38 h g_{i+1}\Big)= y_i+h\Big(
\frac 38 f_{i+1}
+\frac{19}{24}(f_{i}-y_{i}g_{i}) \nonumber \\
&\qquad-\frac{5}{24}(f_{i-1}-y_{i-1}g_{i-1})
+\frac{1}{24}(f_{i-2}-y_{i-2}g_{i-2})
\Big).
\label{e-AM4}
\end{eqnarray}
We again note that the quantities we work with are the spherical
harmonic coefficients of each variable, which are
which are functions of the compactified radius $x_i$, according to
the procedure outlined at the end of Section~\ref{sec:discrete-repr}.

The radial $\Phi$ integration requires special treatment due to the
nonlinear term, $N_{J}$ in \eqref{eq:wev} (given explicitly
in \ref{app:nl}). It is a function of both $J_{,u}$ and $\bar{J}_{,u}$,
and thus, according to \eqref{eq:def-phi}, both $\Phi$ and $\bar{\Phi}$.
We write \eqref{eq:wev} in the form
\begin{equation}
\Phi_{,x}+\Phi a + \bar{\Phi} b =R_\Phi,{}
\label{eq:phi2}
\end{equation}
where
\begin{eqnarray}
a&=& \Big(r(x)\frac{d
x}{d r}\Big)^{-1}+\;\frac{J}{2K}(\bar{J}K_{,x}-\bar{J}_{,x}K)
\nonumber \\
b&=&\frac{J}{2K}(J K_{,x}-J_{,x}K).
\end{eqnarray}
and $R_\Phi$ contains the remaining terms but does not involve $\Phi$ or
$\bar{\Phi}$ or their derivatives.
This is integrated using the scheme of Eq.~(\ref{e-AM4}) to obtain
an equation that involves both $\Phi_{i+1}$ and $\bar{\Phi}_{i+1}$. Taking
the complex conjugate leads to a second equation in the two unknowns, and
solving the system gives
\begin{eqnarray}
\Phi_{i+1} &\bigg(\Big[1+\frac{3h}{8}\bar{a}_{i+1}\Big]
                  \Big[1+\frac{3h}{8}a_{i+1}\Big]
                 -\left(\frac{3h}{8}\right)^2\bar{b}_{i+1} b_{i+1}
                 \bigg)
\nonumber \\
 & =T_i\Big(1+\frac{3h}{8}\bar{a}_{i+1}\Big)-\bar{T}_i
 \frac{3h}{8}b_{i+1},
\end{eqnarray}
where
\begin{eqnarray} \label{eq:rad-int-phi}
T_i=\Phi_i 
&+&\frac{3h}{8}R_{\Phi,i+1}
\nonumber \\
&+&\frac{19h}{24}(R_{\Phi,i}-\Phi_{i}a_{i}-\bar{\Phi}_{i}b_{i})
\nonumber \\
&-&\frac{5h}{24}(R_{\Phi,i-1}-\Phi_{i-1}a_{i-1}-\bar{\Phi}_{i-1}b_{i-1})
\nonumber \\
&+&\frac{h}{24}(R_{\Phi,i-2}-\Phi_{i-2}a_{i-2}-\bar{\Phi}_{i-2}b_{i-2}).
\end{eqnarray}
The scheme above does not allow us to work directly
with the spherical harmonic
coefficients of $\Phi$, $K$, $J$, $K_{,x}$, $J_{,x}$, and $R_\Phi$,{}
due to the non-linear terms $\Phi a$ and $\bar\Phi b$ which must be
evaluated on the collocation grid. To compute \eqref{eq:rad-int-phi},
we therefore
recompose $K$, $J$, $K_{,x}$, $J_{,x}$, and $R_\Phi$ on the collocation grid
defined by \eqref{eq:collocation-points} to perform the required
multiplications. Having evaluated $\Phi$ according to
\eqref{eq:rad-int-phi} on the collocation grid,
we decompose $\Phi$ to obtain its spectral
coefficients in terms of spin $s=2$ real-valued spherical harmonics
${}_{2}Z_{\ell m}$ for each radial point $x_i$.

Note that the radial integration schemes \eqref{eq:rad-scheme} and
\eqref{eq:rad-int-phi} both require data on $3$ radial points to start
the algorithm. These must be provided as boundary data on the
worldtube $\Gamma$.

\subsection{Time integration}
\label{sec:time-int}

The evolution equation for $J$ has the form
\begin{equation} \label{eq:J-time-int}
J_{,u}=\Phi\,.
\end{equation}
This equation can be straightforwardly integrated via a 4th-order Runge-Kutta
scheme using the spectral coefficients of $\Phi$.
In addition, we add numerical dissipation to \eqref{eq:J-time-int}.
To be explicit, we solve
\begin{equation}
(J_i^{\ell m})_{,u}=\Phi_i^{\ell m} + D J_i^{\ell m}\,,{}
 \qquad \forall\; \ell,m,i\,,
\end{equation}
where $D$ is a dissipation operator defined in Section~\ref{sec:rad-deriv}.
Since it is necessary to solve the hypersurface equations to obtain
the $\Phi_i^{\ell m}$, the hypersurface 
equations must be solved for each intermediate Runge-Kutta step.

\subsection{Summary of algorithm}
\begin{enumerate}
 \item 
Assume data for $J$ in the form of spectral coefficients $J^{\ell m}(x_i)$
at (intermediate) timestep $t_n$ for each radial shell $x_i$ .
If $t_n$ is the first timestep, the $J_{\ell m}(x_i)$ are given by initial data.
 \item
Compute $K_{\ell m}(x_i)$, as well as radial derivatives $J_{,x}^{\ell m}(x_i)$
and
$J_{,xx}^{\ell m}(x_i)$ from $J(x_i)$ by means of \eqref{eq:rad-deriv1} and
\eqref{eq:rad-deriv2}, respectively.
Since $K$ is related non-linearly to $J$, we need to recompose $J(x_i)$ from
$J^{\ell m}(x_i)$ to evaluate $K(x_i)$ on the
collocation grid. Afterwards, we decompose $K(x_i)$ to obtain $K^{\ell m}(x_i)$.
 \item
Provide inner worldtube boundary data for $\beta$, $U$, $\hat{W}$ and $\Phi$ at
(intermediate) timestep $t_n$ in terms of spectral coefficients on the first $3$
radial points.
 \item
Integrate hypersurface equations in the order
(i) $\beta$, (ii) $Q$, (iii) $U$, (iv)
$\hat{W}$, and (v) $\Phi$ by using the steps described in
Section~\ref{sec:num-methods} and Section~\ref{sec:rad-int}.
 \item
Evaluate next Runge-Kutta step for $J_{,u}=\Phi$ to
obtain $J$ at next (intermediate) step $t_{n+1}$ as described in
Section~\ref{sec:time-int}.
\end{enumerate}

\subsection{Remarks on the computational implementation}

We have implemented a new code within the \texttt{Cactus}
computational toolkit \cite{Goodale02a, cactusweb}. The underlying
grid array structures are provided by \texttt{Carpet}
\cite{Schnetter-etal-03b,carpeturl}.
Memory handling of the collocation grid, and recomposition/decomposition
in terms of spin-weighted spherical harmonics is provided by
\texttt{SphericalSlice} \cite{ReisswigPhD}.

Since the memory consumption of the implemented code is rather low, and since
the computational efficiency is high, we do not currently decompose the
domain to distribute the work load across multiple processing units.
We do, however, make use of multi-threading via \texttt{OpenMP} to enable
faster processing on shared memory units.
In particular, we use multi-threading in the following two kinds of loops:
(i) when looping over spectral coefficients to perform, for instance, radial
integration for each separate mode, and (ii) when looping over points on the
collocation grid to evaluate non-linear terms in the equations (such as RHS
evaluation). Depending on the number of spectral coefficients and the number of
available cores within one shared memory unit, the observed scaling can be
close to the optimum (though we remark that we have done a rather limited
number of tests on a compute node with up to $12$ shared memory cores).

The implemented code is designed such that it can run concurrently with our
3+1 evolution code \texttt{Llama} \cite{Pollney:2009yz, Pollney:2009ut}.
This is important for future application in on-the-fly Cauchy-characteristic
extraction, where the metric data will be transported to $\scri$ during Cauchy
evolution without the need of an additional post-processing step (which
is currently the
case for the algorithm presented in \cite{Reisswig:2009rx, Babiuc:2010ze}).
Furthermore, this is necessary for a future implementation of Cauchy
characteristic matching\cite{Bishop-etal-1996:Cauchy-characteristic-matching,
Winicour05} in which the characteristic evolution is used to
provide on-the-fly boundary data for a Cauchy evolution.

\section{Results}
\label{sec:results}

\subsection{Linearized solutions}
\label{sec:lin-sol}

To test the convergence of our numerical scheme on a dynamical spacetime,
we use solutions to the linearized Einstein equations in Bondi-Sachs form
on a Minkowski background (Section 4.3 of ~\cite{Bishop-2005b}). We write
\begin{eqnarray}
J^{\rm lin}&=& \sqrt{(\ell -1)\ell(\ell+1)(\ell+2)}\;{}_2Z_{\ell m}
\,\Re(J_\ell(r)\, e^{i\nu
u}), \nonumber \\[0.2\baselineskip]
U^{\rm lin}&=& \sqrt{\ell(\ell+1)}\;{}_1Z_{\ell m} \,\Re(U_\ell(r) \,{}
e^{i\nu u}),
\nonumber \\[0.2\baselineskip]
\beta^{\rm lin}&=& Z_{\ell m} \,\Re(\beta_\ell e^{i\nu u}),
    \nonumber \\[0.2\baselineskip]
\; \hat{W}^{\rm lin}&=& Z_{\ell m}\, \Re(\hat{W}_{\ell}(r) e^{i\nu u}),
\label{e-an}
\end{eqnarray}
where $J_\ell(r)$, $U_\ell(r)$, $\beta_\ell$, $\hat{W}_{\ell}(r)$ are
in general
complex, and taking the real part leads to $\cos(\nu u)$ and $\sin(\nu u)$
terms. The quantities $\beta$ and $\hat{W}$ are real; while $J$ and $U$ are
complex due to the terms $\eth^2 Z_{\ell m}$ and  $\eth Z_{\ell m}$,
representing different terms in the angular part of the metric. We
require a solution that is well-behaved at future null infinity.
We find~\cite{Reisswig:2006}, in the case $\ell=2$,
\begin{eqnarray}
\beta_2&=&\beta_0, \nonumber \\
J_2(r)&=&\frac{24\beta_0 +3 i \nu C_1 - i \nu^3 C_2}{36}+\frac{C_1}{4 r}
       -\frac{C_2}{12 r^3}, \nonumber \\
U_2(r)&=&\frac{-24i\nu \beta_0 +3 \nu^2 C_1 - \nu^4 C_2}{36} +\frac{2\beta_0}{r}
       +\frac{C_1}{2 r^2} +\frac{i\nu C_2}{3 r^3} +\frac{C_2}{4 r^4}, \nonumber \\
\hat{W}_{2}(r)&=&\frac{24i\nu \beta_0 -3 \nu^2 C_1 + \nu^4 C_2}{6} 
           +\frac{3i\nu C_1 -6\beta_0-i\nu^3 C_2}{3r} \nonumber \\
           & &-\frac{\nu^2 C_2}{r^2} +\frac{i\nu C_2}{r^3} +\frac{C_2}{2r^4},
\label{e-NBl2}
\end{eqnarray}
with the (complex) constants $\beta_0$, $C_1$ and $C_2$ freely specifiable; and
in the case $\ell=3$
\begin{eqnarray}
\beta_3&=&\beta_0, \nonumber \\[0.2\baselineskip]
J_3(r)&=&\frac{60\beta_0 +3 i \nu C_1 + \nu^4 C_2}{180}+\frac{C_1}{10 r}
       -\frac{i \nu C_2}{6 r^3} -\frac{C_2}{4r^4},
\nonumber \\[0.2\baselineskip]
U_3(r)&=&\frac{-60i\nu \beta_0 +3 \nu^2 C_1 - i \nu^5 C_2}{180} \nonumber \\[0.2\baselineskip]
      &   &+\frac{2\beta_0}{r} +\frac{C_1}{2 r^2}
           -\frac{2\nu^2 C_2}{3 r^3} +\frac{5 i \nu C_2}{4 r^4}
           + \frac{C_2}{r^5}, \nonumber \\
\hat{W}_{3}(r)&=&\frac{60 i \nu \beta_0 -3 \nu^2 C_1 + i\nu^5 C_2}{15}
  +\frac{i\nu C_1 -2\beta_0+\nu^4 C_2}{3r} \nonumber \\
          & & -\frac{i2\nu^3 C_2}{r^2} -\frac{4i\nu^2 C_2}{r^3}
          +\frac{5\nu C_2}{r^4}+\frac{3 C_2}{r^5}.
\label{e-NBl3}
\end{eqnarray}

We establish convergence by testing the evolution quantities against
linearized solutions listed in Section~\ref{sec:lin-sol}.
The linearized solution provides initial data for $J$ on a null cone,
as well as boundary data for $\beta$, $Q$, $U$,
$\hat{W}$, and $\Phi$ at the worldtube $\Gamma$.
During evolution, we compute the error $\epsilon$ in all evolved quantities
by comparing with the linearized solution.
Since the code solves the general nonlinear case whereas the exact solution
satisfies the linearized Einstein equations, we expect $\epsilon(J)$
to converge towards zero at the order of accuracy of the numerical scheme
only in a regime in which $|\epsilon(J)|$ is much larger than any
nonlinear contribution.

We have performed a number of test cases using $(\ell,m)=(2,2)$ linearized
solutions, $(\ell,m)=(3,3)$ linearized solutions, and a superposition of
both. In the latter case, we compute $J$ via
\begin{equation}
J^{\rm lin} = \sum_{\ell, m} \kappa_\ell\, {}_{2}Z_{\ell m}\,
\Re(J_{\ell}(r) \e^{i \nu u})\,,
\end{equation}
where $\kappa_\ell=\sqrt{(\ell
-1)\ell(\ell+1)(\ell+2)}$.
The remaining superposed linearized solutions for all other quantities  are
constructed in the same way, using the appropriate spin weight for the
${}_{s}Z_{\ell m}$ and the appropriate $\ell$-dependent coefficients
$\kappa_\ell$, respectively (compare \eqref{e-an}).

The linearized solutions depend on free parameters $C_1$, $C_2$, $\beta_0$ and
$\nu$ which we have tested for a range of different values.
Note that the amplitudes $C_1$, $C_2$, and $\beta_0$ must be of linear
order
($\lesssim 10^{-5}$).

In all cases considered, we find better than 4th-order convergence until the
error roughly reaches the square of the amplitude of the linearized
solution, beyond which convergence deteriorates, as expected, due to
the emergence of nonlinear behavior.

\begin{figure*}[t]
\centering
 \includegraphics[width=0.95\textwidth]{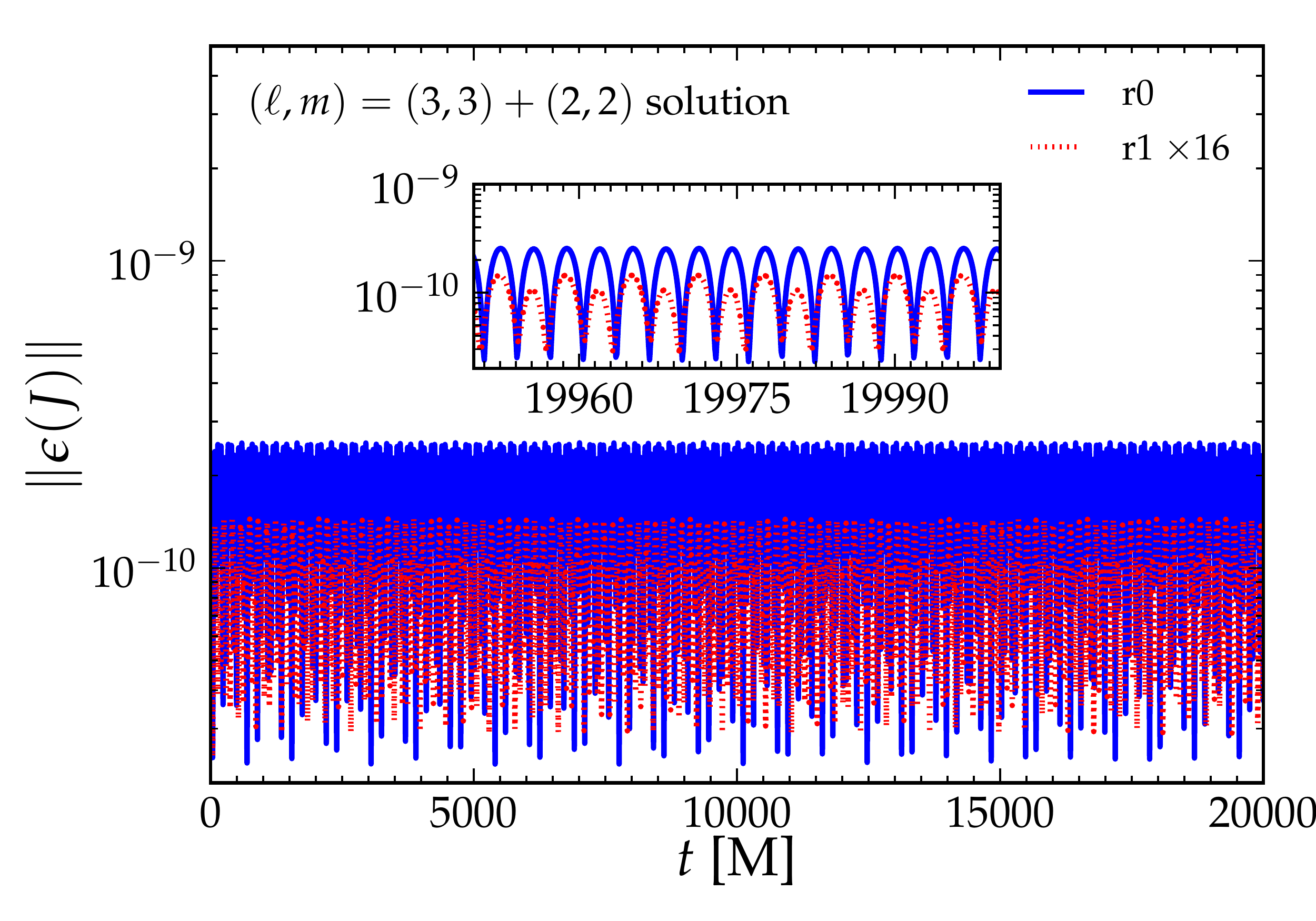}
 \caption{$L_2$ norm \eqref{eq:norm} of the error in the evolution variable $J$
of a superposed $(\ell,m)=(2,2)+(3,3)$ linearized solution on two resolutions
$r0$ and
$r1$ (Table~\ref{tab:res-parameters}). Given that the amplitude of $J$
itself is on the order of $10^{-6}$ (compare \eqref{eq:lin-parameters}), we
observe that the relative error at the given resolutions is $< 10^{-4}$.
The error in the high resolution $r1$ is scaled for 4th-order convergence. We
observe better than 4th-order
convergence (see text for a discussion). The evolution is still stable after
$T=20000M$. This
corresponds to $\sim6400$ cycles in the solution. The inset shows a close up 
of the error for the last $50M$ of evolution.}
  \label{fig:conv_l3l2}
\end{figure*}

As a particular example, we show convergence of a superposed
$(\ell,m)=(2,2)+(3,3)$ solution with parameters
\begin{eqnarray} \label{eq:lin-parameters}
C_1&=3\times10^{-6},\qquad C_2&=1\times10^{-6},\\
\beta_0&=i \times10^{-6},\qquad\quad\nu &= 1.0\,.
\end{eqnarray}
Fig.~\ref{fig:conv_l3l2} plots the $L_2$-norm of the error
$\norm{\epsilon(J)}$
on two resolutions $r0$
and $r1$ (see Table~\ref{tab:res-parameters}) scaled for 4th-order
convergence. We define the $L_2$-norm in terms of the sum over all modes and
radial points by
\begin{equation} \label{eq:norm}
\norm{f} = \sqrt{\sum_{i,\ell,m} \left(f_{\ell m}(x_i)\right)^2}\,.
\end{equation}
The appropriate convergence scaling can be determined from the convergence
rate defined in terms of the grid spacing $\Delta x$,
\begin{equation}
C = \left(\frac{\Delta x_{r0}}{\Delta x_{r1}}\right)^p\,,
\end{equation}
where $p=4$ is the expected order of convergence.
By doubling the resolution, we
expect the higher resolution error, $r1$, to be smaller by a factor of
$16$ given our 4th-order accurate algorithm, i.e.,~we should get
\begin{equation}
C = \left(\frac{\norm{\epsilon(J)}_{r0}}{\norm{\epsilon(J)}_{r1}}\right)^{p=4}
= 16\,.
\end{equation}
As shown in Fig.~\ref{fig:conv_l3l2}, this is
indeed the case. We measure better than 4th-order convergence (see further
below for a discussion).
Furthermore, the evolution is still stable after $T=20000M$
(corresponding to $\sim6400$ cycles of the solution).

The grid settings and parameters for this test are given in
Table~\ref{tab:res-parameters}.
In all cases, we apply radial dissipation of amplitude $\epsilon_{\rm diss}=0.2$.
The inner boundary is located at $R_\Gamma=15M$ (corresponding to
$r_\Gamma=15$, \eqref{e-compactify}). The inner compactified coordinate radius
$x_{\rm in}$ is chosen such that the nominal grid (i.e.,~the grid excluding the
3 inner boundary points) starts at $x_{i=3}=0.36$.

\begin{figure*}[t]
\centering
 \includegraphics[width=0.95\textwidth]{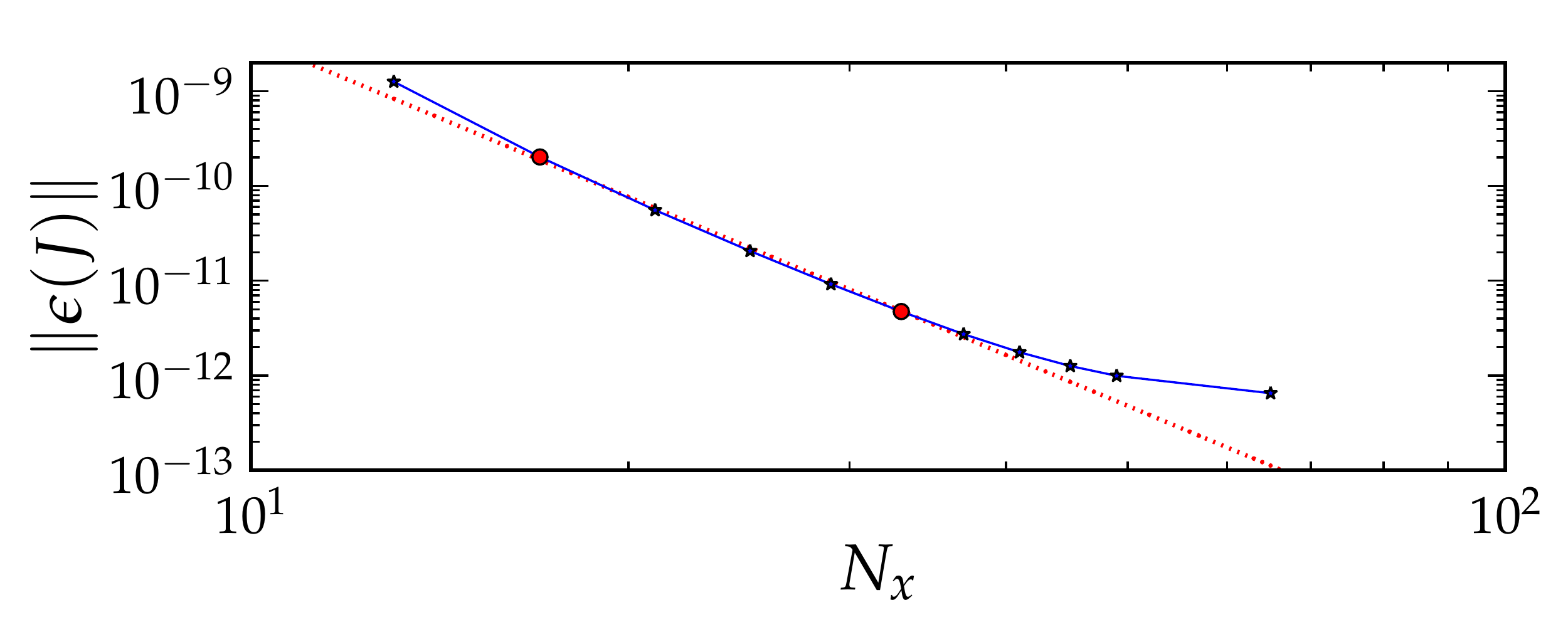}
 \caption{
$L_2$-norm over time \eqref{eq:time-norm} of the error in the evolution
variable $\epsilon(J)$ of a superposed $(\ell,m)=(2,2)+(3,3)$ linearized
solution on multiple radial resolutions. As the radial resolution is increased
starting from $N_x=13$ up to $N_x=65$ points, the error decreases at better than
4th-order. The two highlighted (red) markers correspond to resolutions $r0$ and
$r1$ (Table~\ref{tab:res-parameters}), respectively.
The red dashed line corresponds to a fit excluding the first point and the last
four points. We measure a convergence order of $\sim 5$.
When the error reaches roughly the square of the amplitude of the linearized
solution ($\sim10^{-12}$) where quadratic (and higher order) terms become
important, the
convergence order starts to deteriorate. This is expected since the linearized
solutions only satisfy the Einstein equations to linear order.}
  \label{fig:conv_l3l2_err}
\end{figure*}

In Fig.~\ref{fig:conv_l3l2_err}, we plot the time $L_2$-norm of the
error $\epsilon(J)$, defined in terms of the sum of all modes on all radial
points over all time steps $t_n$
\begin{equation} \label{eq:time-norm}
\norm{f} = \sqrt{\sum_{i,\ell,m,n} \left(f_{\ell m}(x_i,t_n)\right)^2}\,.
\end{equation}
We consider radial resolutions
$N_x=\left[13,17,21,25,29,33,37,41,45,49,65\right]$ with appropriately
adapted time resolutions.
As the resolution is increased, the error drops as expected.
Note that even on the coarsest radial grids, the
code achieves an accuracy with a relative error better
than $\epsilon(J) \approx 10^{-4}$ (the amplitude of our solution is on the
order of $10^{-6}$; compare \eqref{eq:lin-parameters}).
This is a significant improvement in efficiency over
the original 2nd-order scheme developed in
\cite{Bishop97b} including its advancements
\cite{Reisswig:2006, Babiuc:2010ze}.
We also note that when further increasing the radial and time resolution, the
error falls below $\sim 10^{-12}$ and we observe an expected
drop in convergence order due to nonlinear effects.
By fitting a line
through the sampled error norms (red dashed line in
Fig.~\ref{fig:conv_l3l2_err}), we can measure the convergence order. In the
present case, we find $\sim5$. Note that we have excluded the coarsest and the
finest three resolutions from the fit.

We have also checked convergence for each individual hypersurface equation by
systematically setting all other quantities to their linearized solutions, and
have found that while the equations for $\beta$, $Q$, $\Phi$ and $J$ converge
consistently at 4th-order, the equation for $U$ and $\hat{W}$ both achieve
measured convergence orders of $\sim5$ at the given resolutions (the
coefficient of the 5th-order error is apparently larger than the 4th-order
term). These contributions lead to an overall
5th-order convergence for $J$. With sufficiently high resolution
($N_x\gtrsim60$ points) the 5th-order error term in the individual equations
for $U$ and $\hat{W}$ diminishes sufficiently so
that we observe the 4th-order convergence expected of the algorithm.
As we increase the resolution further to the point that the error
approaches the square of the amplitude of the linearized solution,
we start to see the influence of nonlinear terms and the comparison
with the linearized exact solution no longer holds.

\begin{table}[t]
\caption{Numerical grid settings used. $N_x$ denotes the number of radial
points on the nominal compactified grid, $\ell_{\rm max}$ is the number of
angular spectral coefficients used, $N_\theta$ and $N_\phi$ denote the
number of angular collocation points in $\theta$ and $\phi$ direction,
respectively, and $\Delta u$ is the time resolution.
Note that our choice of the number of angular spectral coefficients and
collocation points results in an exact representation of angular derivatives and
functions in the case of a solution with $\ell\leq 3$.}
\label{tab:res-parameters}
\begin{center}
\begin{tabular}{lrrrrr}
\hline
      & $N_x$ & $\ell_{\rm max}$ & $N_\theta$ & $N_\phi$ &
$\Delta u$ \\
\hline
$r0$  & $17$  & $3$              & $8$        & $8$      & $0.1$      \\
$r1$  & $33$  & $3$              & $8$        & $8$      & $0.05$     \\
\hline
\end{tabular}
\end{center}
\end{table}

\subsection{Pseudo random noise}

A strong test for stability involves injecting (pseudo) random noise
into the evolution, via the initial and boundary data~\cite{Szilagyi00a,
Alcubierre2003:mexico-I}. By this method, any exponentially growing error
modes, if present, will be stimulated at a much higher amplitude than
would naturally occur due to truncation or round-off error.

In this test, we add noise to all spectral coefficients of
$J$ on the initial null
hypersurface, and also to all spectral coefficients of all remaining
quantities that are needed at the inner boundary, the worldtube $\Gamma$, at
each timestep. We add noise of amplitude $A_{\rm noise}=10^{-2}$ to
a $(\ell,m)=(3,3)$ linearized solution with parameters given by
\eqref{eq:lin-parameters}. Note that the amplitude of the noise is $10,000$
times stronger than the amplitude of the linearized solution itself, and
thus of nonlinear scale.

\begin{figure*}[t]
\centering
 \includegraphics[width=0.95\textwidth]{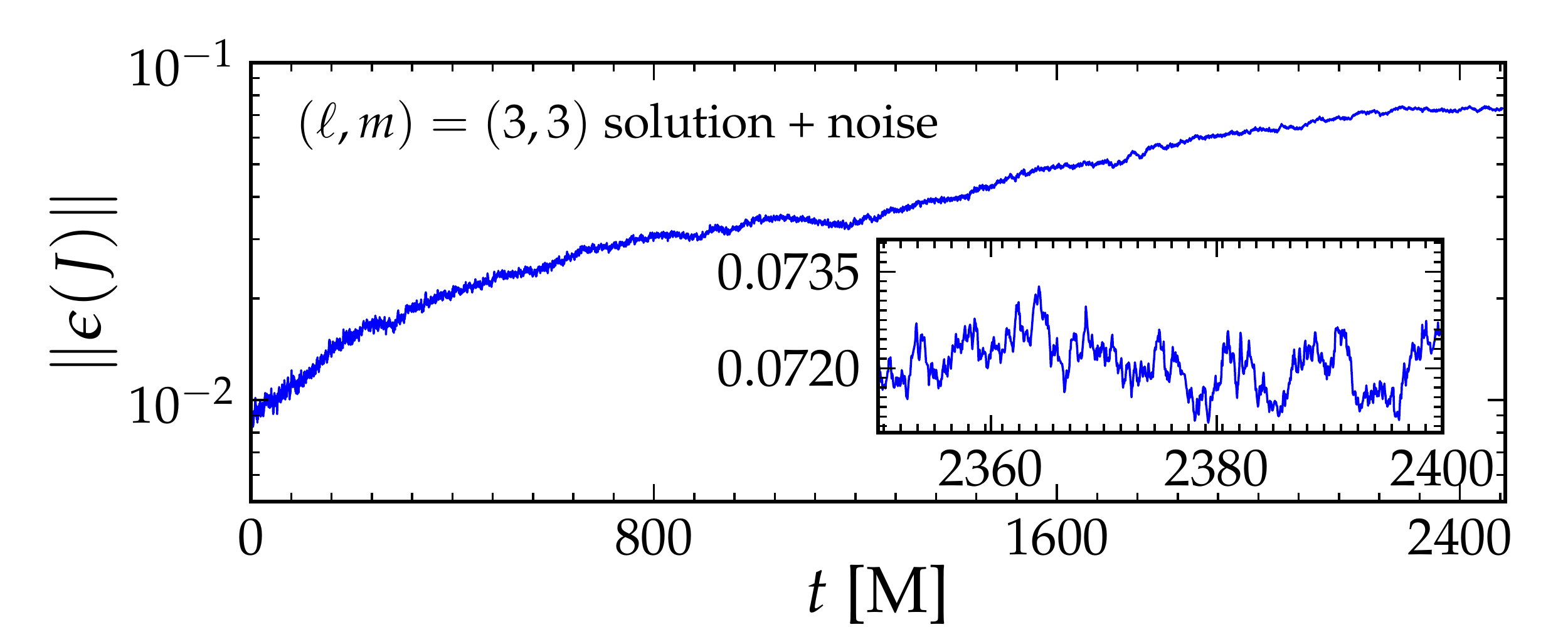}
 \caption{$L_2$ norm \eqref{eq:norm} of the error in the evolution variable $J$
of a $(\ell,m)=(3,3)$ linearized
 solution with pseudo random noise of amplitude $A_{\rm noise}=10^{-2}$ injected
into modes
$\ell \leq 8$ at the  worldtube. The error remains bounded after a few
tens of thousand iterations. This indicates that the scheme is stable even in
the presence of strong noise. The inset shows a close-up of the error over a
timescale of $50M$.}
  \label{fig:noise-l8}
\end{figure*}

In Fig.~\ref{fig:noise-l8}, we show the norm \eqref{eq:norm} of the error
$\epsilon(J)$ over a period of
$T=2500 M$, where we have chosen resolution $r1$ as listed in
Table~\ref{tab:res-parameters}. To allow for higher frequency angular modes,
in the solution we use spectral coefficients up to $\ell=8$, and increase
the number of collocation points to $(N_\theta,N_\phi)=(18,18)$ accordingly.
We inject noise not only into the $(\ell,m)=(3,3)$ solution mode,
but also into all other modes up to $\ell=8$, which would otherwise be zero.
To control the stability\footnote{We have found that lower values of
$\epsilon$ may trigger an instability at $\scri$. This can be cured by either
using larger $\epsilon$ everywhere, or by just increasing $\epsilon$ at
$\scri$.} 
of the scheme, we set the amount of dissipation to $\epsilon_{\rm
diss}=0.5$.
The test demonstrates that the non-linear coupling of angular modes does not
lead to unstable behavior over the observed timescale.

\section{Discussion}
\label{sec:discussion}

We have developed and implemented a high-order algorithm for numerically
integrating the full non-linear Einstein equations along characteristic
null hypersurfaces in the Bondi-Sachs framework. 
The implemented code evolves the full Einstein equations in the wave
zone of a compact body, remaining stable, convergent, and achieving high
accuracy with relatively low computational cost compared to the previous
2nd-order algorithm.
Radial integration is
performed in terms of a modified Adams-Moulton scheme that is 4th-order
accurate. Radial derivatives are computed in terms of 4th-order finite
differences. Angular derivatives are computed in terms of spectral
expansions of real-valued spin-weighted spherical harmonics. 
It is in principle straightforward to design an algorithm along the lines used
here at even higher order of convergence than 4th-order. However, the
stability properties of the algorithm are clearly dependent on its order,
and achieving stability for an algorithm of higher order may be problematic.

The
implemented algorithm is a first step towards more
efficient gravitational-wave extraction algorithms via
Cauchy-characteristic extraction. It is also a first step towards Cauchy-characteristic matching in which the characteristic evolution
is used to
provide on-the-fly boundary data for a 3+1 evolution. In view of
this, we have designed our code such that it can run concurrently with our
3+1 evolution code. The next step will be the implementation of an
improved algorithm for worldtube boundary data transformation that couple
Cauchy and characteristic evolutions, and that is of higher than
2nd-order.

\ack

The authors would like to thank Peter Diener for providing dissipation operator
stencil coefficients, and Harald Pfeiffer for comments on the manuscript. We
thank the Erwin Schroedinger Institute, Austria, Universitas de les Illes
Balears, Spain,
and Rhodes University, South Africa, for hospitality. 
This work is supported by
the National Science Foundation under grant numbers AST-0855535 and
OCI-0905046. CR acknowledges support by NASA through Einstein Postdoctoral
Fellowship grant number PF2-130099 awarded by the Chandra X-ray center, which is
operated by the Smithsonian Astrophysical Observatory for NASA under contract
NAS8-03060.
NTB has been supported by the National Research Foundation,
South Africa. 
Computations were performed on 
the LONI network (\texttt{www.loni.org}) under allocation
\texttt{loni\_numrel06} and \texttt{loni\_numrel07}, 
and the Caltech compute 
cluster ``Zwicky'' (NSF MRI award No.\ PHY-0960291).


\appendix

\section{The nonlinear terms in the Einstein equations}
\label{app:nl}

The nonlinear terms $N_\beta, N_Q, N_U, N_W$ and $N_J$ in \eqref{eq:beta} through
\eqref{eq:wev} were first presented in~\cite{Bishop97b}. We repeat
them here,
but with a mis-print in Eq.~(A3) of~\cite{Bishop97b} corrected.
\begin{equation}
N_\beta=\frac{r}{8}\left(J_{,r}\bar J_{,r}-K^2_{,r} \right).
\label{eq:nbeta}
\end{equation}
\begin{equation}
N_U=\frac{e^{2\beta}}{r^2} \left(KQ-Q-J\bar Q \right),
\label{eq:nu}
\end{equation}
\begin{eqnarray}
N_Q &=& r^2 \Bigg( (1-K) ( \eth K_{,r} + \bar \eth J_{,r} ) + \eth (\bar J
J_{,r} ) + \bar \eth ( J K_{,r} )  - J_{,r} \bar \eth K \nonumber \\
   & + & \frac{1}{2K^2}(\eth \bar J (J_{,r} - J^2 \bar J_{,r} ) + \eth J (\bar
J_{,r} -\bar J^2  J_{,r} ) ) \Bigg).
\label{eq:nq}
\end{eqnarray}
\begin{eqnarray}
N_W&=& e^{2 \beta} \Bigg( (1-K) ( \eth \bar \eth \beta +  \eth \beta \bar \eth
\beta) + \frac{1}{2} \bigg ( J (\bar \eth \beta)^2 + \bar J (\eth \beta)^2
\bigg) \nonumber \\
& & - \frac{1}{2} \bigg ( \eth \beta ( \bar \eth K - \eth \bar J) + \bar \eth
\beta ( \eth K - \bar \eth J ) \bigg) + \frac{1}{2} \bigg ( J \bar \eth^2 \beta
+ \bar J \eth^2 \beta  \bigg) \Bigg )  \nonumber \\
& & - e^{-2 \beta} \frac{r^4}{8} ( 2 K U_{,r} \bar U_{,r} + J \bar U^2_{,r} +
\bar J U^2_{,r}).
\label{eq:mw}
\end{eqnarray}
\begin{equation}
N_J=N_{J1}+N_{J2}+N_{J3}+N_{J4}+N_{J5}+N_{J6}+N_{J7}+\frac{J}{r}(P_1+P_2
+P_3+P_4)
\label{eq:nj}
\end{equation}
where
\begin{eqnarray}
N_{J1}&=& - \frac{e^{2 \beta}}{r} \bigg ( K ( \eth J \bar \eth \beta + 2 \eth K
\eth \beta - \bar \eth J \eth \beta) + J ( \bar \eth J \bar \eth \beta - 2 \eth
K \bar \eth \beta) - \bar J \eth J \eth \beta \bigg) ,
\nonumber \\
N_{J2}&=& -\frac{1}{2} \bigg ( \eth J ( r \bar U_{,r} + 2 \bar U) + \bar \eth J
( r U_{,r} + 2 U) \bigg) ,
\nonumber \\
N_{J3}&=& (1-K) ( r  \eth U_{,r} + 2 \eth U) - J ( r \eth \bar U_{,r} + 2 \eth
\bar U) ,
\nonumber \\
N_{J4}&=& \frac{r^3}{2} e^{-2 \beta} \bigg( K^2 U^2_{,r} + 2 J K U_{,r} \bar
U_{,r} + J^2 \bar U^2_{,r} \bigg) ,
\nonumber \\
N_{J5}&=& - \frac{r}{2} J_{,r} ( \eth \bar U + \bar \eth U) ,
\nonumber \\
N_{J6}&=& r \Bigg( \frac{1}{2} ( \bar U \eth J + U \bar \eth J ) (J \bar J_{,r}
- \bar J J_{,r} )  \nonumber \\
    & & + ( J K_{,r} - K J_{,r} ) \bar U \bar \eth J
     - \bar U ( \eth J_{,r} - 2 K \eth K J_{,r} + 2 J \eth K K_{,r} ) \nonumber
\\
    & & - U ( \bar \eth J_{,r} - K \eth \bar J J_{,r} + J \eth \bar J  K_{,r} )
\Bigg) ,
\nonumber \\
N_{J7}&=& r ( J_{,r} K -  J K_{,r} ) \bigg ( \bar U ( \bar \eth J - \eth K ) +
U ( \bar \eth K - \eth \bar J ) \nonumber \\
       & & + K ( \bar \eth U - \eth \bar U ) + ( J \bar \eth \bar U - \bar J
\eth U ) \bigg) ,
\nonumber \\
P_1 &=& r^2 \bigg ( \frac{J_{,u}}{K} (\bar J_{,r} K - \bar J K_{,r} ) +
\frac{\bar J_{,u}}{K} ( J_{,r} K -  J K_{,r} ) \bigg)
	    - 8 \left(r+r^2\hat{W}\right) \beta_{,r} ,
\nonumber \\
P_2 &=& e^{2 \beta} \Bigg( - 2 K ( \eth \bar \eth \beta + \bar \eth \beta \eth
\beta) - ( \bar \eth \beta \eth K
                     + \eth \beta  \bar \eth K) \nonumber \\
                & + &   \bigg( J ( \bar \eth^2 \beta + (\bar \eth \beta)^2 ) +
                           \bar J ( \eth^2 \beta + (\eth \beta)^2 ) \bigg)
               + ( \bar \eth J \bar \eth \beta + \eth \bar J \eth \beta) \Bigg),
\nonumber \\
P_3 &=& \frac{r}{2} \bigg( ( r \bar \eth U_{,r} + 2 \bar \eth U) +
		       ( r  \eth \bar U_{,r} + 2 \eth \bar U) \bigg) ,
\nonumber \\
P_4 &=& - \frac{r^4}{4} e^{-2 \beta} ( 2 K U_{,r} \bar U_{,r} + J \bar U^2_{,r}
		      + \bar J U^2_{,r} ).
\label{eq:nji}
\end{eqnarray}

\section{Regularized equations at $\scri$}
\label{app:reg-scri}

At $\scri$, the equations simplify when inserting \eqref{e-compactify}
and \eqref{eq:jac} and taking the limit $r\rightarrow\infty$.
In particular, terms containing powers of $1/r,...,1/r^n$ vanish.
Below, we give the equations evaluated at $\scri$.
\begin{eqnarray}
    \beta_{,x} &=& 0\,,                            \label{eq:beta-scri} \\
    Q &=& -2\eth\beta\,, \label{eq:wq-scri} \\
    U_{,x} &=& r_\Gamma^{-1}e^{2\beta}(KQ-J\bar{Q})\,,               
\label{eq:wua-scri} \\
    \hat{W} &=& \frac{\eth\bar{U}+\bar\eth U}{2}\,, \label{eq:ww-scri} \\
    \Phi &=& -\eth U + N_{J,\scri}\,,
\end{eqnarray}
where
\begin{equation}
N_{J,\scri} = - \frac{\bar{U} \eth J + U \bar{\eth}J}{2} + (1-K)\eth U -
\frac{J(\eth\bar{U}-\bar{\eth}U)}{2}\,.
\end{equation}

\section{Dissipation operator stencil}
\label{app:diss}

Close to the outer boundary where we do not have enough points to apply
the standard centred-stencil Kreiss-Oliger dissipation operator
\eqref{eq:diss},
we use side-winded dissipation operator
stencils derived for SBP operator $D_{4-2}$ of \cite{Diener05b1}.
This particular dissipation operator
is defined via coefficients $a_{ij}$ and
$q_{i}$.
In the interior of the domain the operator reads
\begin{eqnarray}
  A_{ij} u_j & = & \frac{\epsilon_{\rm diss}}{2^{2p}} \left[ q_0 u_i +
    \sum_{j=1}^7 q_{j} \left( u_{i-j} + u_{i+j} \right) \right]
\end{eqnarray} 
where the $q_i$ are the coefficients given in \eqref{eq:diss}.
Near the outer boundary (on points $i=N_x-5,...,N_x-1$), the operator can be
written as
\begin{eqnarray}
  A_{ij} u_j & = & \frac{\epsilon_{\rm diss}}{2^{2p}} \sum_{j=1}^{7}
a_{j,N_x-i} u_{N_x-j}, 
\end{eqnarray}
where the coefficients $a_{ij}$ are taken from \cite{Diener05b1} reported below.
We note that $\epsilon_{\rm diss}\ge 0$ selects the amount of dissipation and is
usually of
order unity.

\begin{equation}
\renewcommand{\arraystretch}{1.5}
(a_{ij}) = 
\left(\begin{array}{ccccccc}
-\frac{48}{17} & \frac{144}{17} & -\frac{144}{17} & \frac{48}{17} & 0 & 0 & 0 \\
   \frac{144}{59} & -\frac{480}{59} & \frac{576}{59} & -\frac{288}{59} &
\frac{48}{59} & 0 & 0 \\
   \frac{144}{43} & \frac{576}{43} & -\frac{912}{43} & \frac{720}{43} &
-\frac{288}{43} & \frac{48}{43} & 0 \\
   \frac{48}{49} & -\frac{288}{49} & \frac{720}{49} & -\frac{960}{49} &
\frac{720}{49} & -\frac{288}{49} & \frac{48}{49}
\end{array}\right) 
\end{equation}


\section*{References}
\bibliographystyle{unsrt}
\bibliography{references,references2}

\end{document}